\documentclass[11pt]{revtex4}
\usepackage{amssymb,epsf}
\usepackage{latexsym}
\begin{document}

\title{Thermodynamics of black holes in $
(n+1)$-dimensional Einstein-Born-Infeld-dilaton gravity}
\author{A. Sheykhi $^{1,2}$\footnote{email address:
asheykhi@mail.uk.ac.ir} and N. Riazi $^{1}$ \footnote{email:
riazi@physics.susc.ac.ir}}
\address{$1$. Physics Department and Biruni Observatory, Shiraz University, Shiraz 71454, Iran\\
         $2$. Physics Department, Shahid Bahonar University, Kerman, Iran}

\begin{abstract}
We construct a new class of $(n+1)$-dimensional $(n\geq3)$ black
hole solutions in Einstein-Born-Infeld-dilaton gravity with
Liouville-type potential for the dilaton field and investigate
their properties. These solutions are neither asymptotically flat
nor (anti)-de Sitter. We find that these solutions can represent
black holes, with inner and outer event horizons, an extreme black
hole or a naked singularity provided the parameters of the
solutions are chosen suitably. We compute the thermodynamic
quantities of the black hole solutions and find that these
quantities satisfy the first law of thermodynamics. We also
perform stability analysis and investigate the effect of dilaton
on the stability of the solutions.
\end{abstract}
\maketitle
\section{Introduction}
Born-Infeld electrodynamics was first introduced in 1934 in order
to obtain a classical theory of charged particles with finite
self-energy \cite{BI}. In this regard the theory was successful.
However, they did not remove all of the singularities associated
with a point charge and nonlinear electrodynamics became less
popular with the introduction of QED which provided much better
agreement with experiment. In recent years,  there has been great
interest in the Born-Infeld type generalizations of Abelian and
non-Abelian gauge theories. Such generalizations arise naturally
in open superstrings and in D-branes \cite{Lei}. The low energy
effective action for an open superstring in loop calculations
leads to Born-Infeld type actions \cite{Frak}. It has also been
observed that the Born-Infeld action arises as an effective action
governing the dynamics of vector-fields on D-branes \cite{Tse}.
The nonlinearity of the electromagnetic field brings remarkable
properties to avoid the black hole singularity problem which may
contradict the strong version of the Penrose cosmic censorship
conjecture in some cases. Actually a new nonlinear
electromagnetism was proposed, which produces a nonsingular exact
black hole solution satisfying the weak energy condition
\cite{Sal}, and has distinct properties from Bardeen black holes
\cite{Bord}. Unfortunately, the Born-Infeld model is not of this
type. The Born-Infeld action including a dilaton and an axion
field, appears in the coupling of an open superstring and an
Abelian gauge field theory. This action, describing a
Born-Infeld-dilaton-axion system coupled to Einstein gravity, can
be considered as a non-linear extension in the Abelian field of
Einstein-Maxwell-dilaton-axion gravity. Some efforts have been
done to construct exact solutions of Einstein-Born-Infeld (EBI)
gravity. Exact solutions of Born-Infeld systems with zero
cosmological constant has been studied by many authors
\cite{Gar,Wil}. Thermodynamics of Born-Infeld black holes have
been discussed in \cite{Rash}. The exact five-dimensional charged
black hole solution in Lovelock gravity coupled to Born-Infeld
electrodynamics was presented in \cite{Aie}. Black hole solutions
to the theory with derivative corrections to the Born-Infeld
action were derived in \cite{Tamaki}. A class of slowly rotating
black hole solutions in Born-Infeld theory has been obtained in
\cite{Diego}. In the presence of cosmological constant, black hole
solutions of Einstein-Born-Infeld (EBI) theory with positive
\cite{Fern1,Dey}, zero or negative constant curvature horizons
have also been constructed \cite{Cai2}. Recently, the
thermodynamical properties of these black hole solutions in (A)dS
spacetime have been studied \cite{Fern2}. Unfortunately, exact
solutions to the Einstein-Born-Infeld equation coupled to matter
fields are too complicated to find except in a limited number of
cases. Indeed, exact solutions to the Einstein Born-Infeld dilaton
(EBId) gravity are known only in three dimensions \cite{YI}.
Numerical studies of the EBId system in four dimensional static
and spherically symmetric spacetime have been done in \cite{Tam}.
Some limited classes (factorized solutions) of four dimensional
black hole in EBId theory with magnetic or electric charge have
also been constructed \cite{yaz,SRM}. Till now, exact black hole
solutions of EBId gravity in more than four dimensions have not
been constructed. Our aim in this paper is to construct a new
class of exact, spherically symmetric solution in
$(n+1)$-dimensional Einstein-Born-Infeld-dilaton gravity for an
arbitrary value of coupling constant with one and two
Liouville-type potentials and investigate their properties.
Specially, we want to perform a stability analysis and investigate
the effect of dilaton on the stability of the solutions.

The organization of this paper is as follows: In Sec. \ref{Field},
we construct a class of $(n+1)$-dimensional black hole solutions
in EBId theory with one and two liouville type potentials and
general dilaton coupling and investigate their properties. In Sec.
\ref{Therm}, we obtain the conserved and thermodynamical
quantities of the $(n+1)$-dimensional black hole solutions and
show that these quantities satisfy the first law of
thermodynamics. In Sec. \ref{stab}, we perform a stability
analysis and show that the dilaton creates an unstable phase for
the solutions. We finish our paper with some concluding remarks.

\section{Field Equations and Solutions\label{Field}}
We consider the $(n+1)$-dimensional $(n\geq3)$ action in which
gravity is coupled to dilaton and Born-Infeld fields
\begin{equation}\label{Act}
S=\int{d^{n+1}x\sqrt{-g}\left(\mathcal{R}\text{
}-\frac{4}{n-1}(\nabla \Phi )^{2}-V(\Phi )+L(F,\Phi)\right)},
\end{equation}
where $\mathcal{R}$ is the Ricci scalar curvature, $\Phi $ is the
dilaton field and $V(\Phi )$ is a potential for $\Phi $. The
Born-Infeld $L(F,\Phi)$ part of the action is given by
\begin{equation}
L(F,\Phi)=2\gamma e^{\frac{4\alpha \Phi }{n-1}}\left( 1-\sqrt{1+\frac{e^{-\frac{%
8\alpha \Phi }{n-1}}F^2}{\gamma}}\right),
\end{equation}
Here, $\alpha $ is a constant determining the strength of coupling
of the scalar and electromagnetic field  and $F^2=F_{\mu \nu
}F^{\mu \nu }$, where $F_{\mu \nu }=\partial _{\mu }A_{\nu
}-\partial _{\nu }A_{\mu }$ is the electromagnetic field tensor
and $A_{\mu }$ is the electromagnetic vector potential. $\gamma $
is called the Born-Infeld parameter with dimension of mass. In the
limit $\gamma \rightarrow \infty $, $L(F,\Phi)$ reduces to the
standard Maxwell field coupled to a dilaton field
\begin{equation}
L(F,\Phi)=-e^{-\frac{4\alpha \Phi }{n-1}}F_{\mu \nu }F^{\mu \nu }.
\end{equation}
On the other hand, $L(F,\Phi)\rightarrow 0$ as $\gamma \rightarrow
0$. It is convenient to set
\begin{equation}
L(F,\Phi)=2\gamma e^{\frac{4\alpha \Phi }{n-1}}{\mathcal{L}}(Y),
\end{equation}
where
\begin{eqnarray}
{\mathcal{L}}(Y) &=&1-\sqrt{1+Y},\label{LY}\\
Y&=& \frac{e^{-\frac{8\alpha \Phi }{n-1}}F^2}{\gamma}.\label{Y}
\end{eqnarray}

The equations of motion can be obtained by varying the action
(\ref{Act}) with respect to the gravitational field $g_{\mu \nu
}$, the dilaton field $\Phi $ and the gauge field $A_{\mu }$ which
yields the following field equations
\begin{equation}\label{FE1}
{\cal R}_{\mu\nu} = \frac{4}{n-1}\left( \partial _{\mu }\Phi
\partial _{\nu }\Phi +\frac{1}{4}g_{\mu \nu }V(\Phi )\right)-
4e^{\frac{-4\alpha \Phi }{n-1}}\partial_{Y}{{\cal L}}(Y)
F_{\mu\eta} F_{\nu}^{\text{ }\eta }+\frac{2\gamma}{n-1}
e^{\frac{4\alpha \Phi }{n-1}} \left[2Y\partial_{Y}{{\cal
L}}(Y)-{{\cal L}}(Y)\right]g_{\mu\nu},
\end{equation}
\begin{equation}\label{FE2}
\nabla ^{2}\Phi =\frac{n-1}{8}\frac{\partial V}{\partial \Phi}+
\gamma\alpha  e^{\frac{4\alpha \Phi }{n-1}}\left[2{
Y}\partial_{Y}{{\cal L}}(Y)-{\cal L}(Y)\right],
\end{equation}
\begin{equation}\label{FE3}
\partial _{\mu }\left( \sqrt{-g}e^{\frac{-4\alpha \Phi }{n-1}}
\partial_{Y}{{\cal L}}(Y) F^{\mu\nu}\right)=0.
\end{equation}
In particular, in the case of the linear electrodynamics with
${\cal L}(Y)=-{1\over 2}Y$, the system of equations
(\ref{FE1})-(\ref{FE3}) reduce to the well-known equations of
Einstein-Maxwell dilaton gravity \cite{CHM}.

We wish to find static and spherically symmetric solutions of the
above field equations. The most general such metric can be written
in the form
\begin{equation}\label{metric}
ds^2=-f(r)dt^2 + {dr^2\over f(r)}+ r^2R^2(r)d\Omega^2_{n-1},
\end{equation}
where, $d\Omega^2_{n-1}$ denotes the metric of an unit $(n-1)$
sphere. The functions $f(r)$ and $R(r)$ should be determined. The
electromagnetic fields equation can be integrated immediately,
where all the components of $F_{\mu\nu}$ are zero except $
F_{tr}$:
\begin{equation}\label{Ftr}
F_{tr}=\frac{q e^{\frac{4\alpha \Phi }{n-1}}}{\left( rR\right)
^{n-1} \sqrt{1+{\frac{2q^{2}}{\gamma \left( rR\right) ^{2n-2}}}}},
\end{equation}
where $q$ is an integration constant related to the electric
charge of the black hole. It is interesting to consider three
limits of (\ref{Ftr}). First, for large $\gamma$ (where the BI
action reduces to Maxwell case) we have $F_{tr}=\frac{q
e^{2\alpha\Phi}}{(rR)^{n-1}}$ as presented in \cite{CHM}. On the
other hand, if $\gamma\rightarrow 0$ we get $F_{tr}=0$. Finally,
in the absence of the dilaton field ($\alpha=0$), it reduces to
the case of $(n+1)$-dimensional Einstein-Born-Infeld theory (see
for example \cite{Dey,Cai2}).

\subsection{ Black hole solutions with a Liouville type
potential }\label{1V}

First, we consider the action (\ref{Act}) with a Liouville type
potential,
\begin{equation}\label{v1}
V(\Phi)=2\Lambda e^{2\zeta\Phi},
\end{equation}
where $\Lambda$ and $\zeta$ are constants. One may refer to
$\Lambda $ as the cosmological constant, since in the absence of
the dilaton field the action reduces to the action of EBI gravity
with cosmological constant\cite{Dey,Cai2}. Here, we have redefined
the cosmological constant as $\Lambda = -n(n-1)/2l^2$. In order to
solve the system of equations (\ref{FE1}) and (\ref{FE2}) for
three unknown functions $f(r)$, $R(r)$ and $\Phi (r)$, we make the
ansatz
\begin{equation}
R(r)=e^{2\alpha \Phi /(n-1)}.\label{Rphi}
\end{equation}

Using (\ref{Rphi}), the electromagnetic fields (\ref{Ftr}) and the
metric (\ref{metric}), one can easily show that equations
(\ref{FE1}) and (\ref{FE2}) have solutions of the form
\begin{eqnarray}
f(r)&=&-{\frac { \left(n-2 \right)\left( { \alpha}^{2}+1 \right)
^{2}{b}^{-2\beta}{r}^{2\beta}}{\left( {
\alpha}^{2}-1 \right)  \left(n+{\alpha}^{2}-2 \right) }}-\frac{m}{%
r^{(n-1)(1-\beta
)-1}}+\frac{2\gamma(\alpha^{2}+1)^{2}{b}^{2\beta}{r
}^{2(1-\beta)}}{(n-1)(n-\alpha^{2})}  \nonumber \\
&&-\frac{2(\alpha ^{2}+1)b^{(3-n)\beta}}{n-1}r^{(n-1)(\beta-1)+1}\int {%
\Gamma r^{-2\beta }dr},  \label{f1}
\end{eqnarray}
\begin{equation}\label{phi}
\Phi (r)=\frac{(n-1)\alpha }{2(1+\alpha ^{2})}\ln (\frac{b}{r}),
\label{phi}
\end{equation}
where
\begin{equation}
\Gamma \equiv \sqrt{\gamma \left( 2q^{2}+\gamma b^{2\beta
(n-1)}r^{2(n-1)(1-\beta )}\right)}.  \label{Gamma}
\end{equation}
The integral can be done in terms of hypergeometric functions and
can be written in a compact form. The result is
\begin{eqnarray}
f(r) &=&-{\frac { \left(n-2 \right)\left( { \alpha}^{2}+1 \right)
^{2}{b}^{-2\beta}{r}^{2\beta}}{\left( { \alpha}^{2}-1 \right)
\left(n+{\alpha}^{2}-2 \right) }}-\frac{m}{r^{(n-1)(1-\beta
)-1}}-2\frac{\gamma (\alpha
^{2}+1)^{2}b^{2\beta }r^{2(1-\beta )}}{(n-1)(\alpha ^{2}-n)}\times  \nonumber \\
&&\left( 1-\text{{\ }}_{2}F_{1}\left( \left[ -\frac{1}{2},\frac{\alpha ^{2}-n%
}{2n-2}\right] ,\left[ \frac{\alpha ^{2}+n-2}{2n-2}\right] ,\frac{-2q^{2}}{%
\gamma b^{2\beta (n-1)}r^{2(n-1)(1-\beta )}}\right) \right),
\end{eqnarray}
Here $b$ is a arbitrary constant and $\beta =\alpha ^{2}/(\alpha
^{2}+1)$. In the above expression, $m$ appears as an integration
constant and is related to the ADM mass of the black hole. In
order to fully satisfy the system of equations, we must have
\begin{eqnarray}\label{lam}
\zeta &=&\frac{2}{\alpha(n-1)},\nonumber\\
 \Lambda &=& \frac{(n-1)(n-2)\alpha^2 }{2b^2(\alpha^2-1)}.
\end{eqnarray}
One may note that as $\gamma \longrightarrow \infty $ these
solutions reduce to the $(n+1)$-dimensional charged dilaton black
hole solutions given in Ref. \cite{CHM}. In the absence of a
non-trivial dilaton ($\alpha =\beta =0 $), the above solutions
reproduce correctly the asymptotically flat Born-Infeld black hole
(see for example \cite{Rash}). Using the fact that
$_2F_1(a,b,c,z)$ has a convergent series expansion for $|z| <1$,
we can find the behavior of the metric for large $r$. This is
given by
\begin{eqnarray}
f(r) &=&-{\frac { \left(n-2 \right)\left( { \alpha}^{2}+1 \right)
^{2}{b}^{-2\beta}{r}^{2\beta}}{\left( { \alpha}^{2}-1 \right)
\left(n+{\alpha}^{2}-2 \right) }}-\frac{m}{r^{(n-1)(1-\beta
)-1}}+\frac{2(\alpha ^{2}+1)^{2}b^{-2(n-2)\beta
}q^{2}}{(n-1)(\alpha ^{2}+n-2)r^{2(n-2)(1-\beta )}}
\nonumber \\
&&-\frac{(\alpha ^{2}+1)^{2}b^{-2(2n-3)\beta }q^{4}}{\gamma
(n-1)(\alpha ^{2}+3n-4)r^{2(2n-3)(1-\beta )}}.
\end{eqnarray}
Note that in the limit $\gamma \rightarrow \infty$ and $\alpha
=\beta= 0$, it has the form of Reissner-N\"{o}rdstrom black hole.
The last term in the right hand side of the above expression is
the leading Born-Infeld correction to the RN black hole with
dilaton field in the large $\gamma$ limit.

\subsection{ Black hole solutions with two Liouville type
potential }\label{2V}

Second, we present exact, $(n+1)$-dimensional solutions to the
EBId gravity equations with an arbitrary dilaton coupling
parameter $\alpha$ and dilaton potential
\begin{equation}\label{v2}
V(\Phi) = 2\Lambda_{0} e^{2\zeta_{0}\Phi} +2 \Lambda e^{2\zeta
\Phi}.
\end{equation}
where $\Lambda_{0}$,  $\Lambda$, $ \zeta_{0}$ and $ \zeta$ are
constants. This kind of  potential was previously investigated by
a number of authors both in the context of $FRW$ scalar field
cosmologies \cite{ozer} and EMd black holes \cite{CHM,yaz2,SR}.
This generalizes further the potential (\ref{v1}). If
$\zeta_{0}=\zeta$, then (\ref{v2}) reduces to (\ref{v1}), so we
will not repeat these solutions, thus we require $\zeta_{0}\neq
\zeta$. Again, using (\ref{Rphi}), the electromagnetic fields
(\ref{Ftr}) and the metric (\ref{metric}), one can easily show
that equations (\ref{FE1}) and (\ref{FE2}) have solutions of the
form
\begin{eqnarray}
f(r)&=&-{\frac { \left(n-2 \right)\left( { \alpha}^{2}+1 \right)
^{2}{b}^{-2\beta}{r}^{2\beta}}{\left( {
\alpha}^{2}-1 \right)  \left(n+{\alpha}^{2}-2 \right) }}-\frac{m}{%
r^{(n-1)(1-\beta )-1}}+\frac{2\Lambda \left( {\alpha}^{2}+1
\right) ^{2}{b}^{2 \beta}r^{2(1-\beta)}}{(n-1)(\alpha^{2}-n
)}\nonumber \\
&&+{\frac {2\gamma\ \left( {\alpha}^{2}+1 \right)
^{2}{b}^{2\,\beta}{r
}^{2(1-\beta)}}{(n-1)(n-{\alpha}^{2})}}-\frac{2(\alpha
^{2}+1)b^{(3-n)\beta }}{n-1} r^{(n-1)(\beta-1)+1}\int {\Gamma
r^{-2\beta }dr},  \label{f2}
\end{eqnarray}
where $b$ is a arbitrary constant and  $m$ the mass parameter.
$\Phi(r)$ and $\Gamma$ are given by Eqs.
(\ref{phi})-(\ref{Gamma}). Setting $\Lambda=0$ in (\ref{f2}) one
recover (\ref{f1}). In order to fully satisfy the system of
equations, we should have $\zeta=2\alpha/(n-1)$, and a similar
relation to Eq. (\ref{lam}) for $\zeta_{0}$ and $\Lambda_{0}$. We
can also turn out the integration and express the solution in
terms of hypergeometric function
\begin{eqnarray}
f(r) &=&-{\frac { \left(n-2 \right)\left( { \alpha}^{2}+1 \right)
^{2}{b}^{-2\beta}}{\left( { \alpha}^{2}-1 \right)
\left(n+{\alpha}^{2}-2 \right)
}}{r}^{2\beta}-\frac{m}{r^{(n-1)(1-\beta )-1}}+\frac{2\Lambda
\left( {\alpha}^{2}+1 \right) ^{2}{b}^{2
\beta}}{(n-1)(\alpha^{2}-n )}r^{2(1-\beta)}-\frac{2\gamma (\alpha
^{2}+1)^{2}b^{2\beta }r^{2(1-\beta )}}{(n-1)(\alpha ^{2}-n)}
\nonumber
\\
&&\times
\left( 1-\text{{\ }}_{2}F_{1}\left( \left[ -\frac{1}{2},\frac{\alpha ^{2}-n%
}{2n-2}\right] ,\left[ \frac{\alpha ^{2}+n-2}{2n-2}\right] ,\frac{-2q^{2}}{%
\gamma b^{2\beta (n-1)}r^{2(n-1)(1-\beta )}}\right) \right)
\end{eqnarray}

One may note that as $\gamma \longrightarrow \infty $ these
solutions reduce to the $(n+1)$-dimensional charged dilaton black
hole solutions given in Ref. \cite{CHM}. In the absence of a
non-trivial dilaton ($\alpha =\beta =0 $), the above solutions
reduce to the Born-Infeld black hole in (A)dS space presented in
\cite{Dey,Cai2}. Again, we can find the behavior of the metric for
large $r$. This is given by
\begin{eqnarray}
f(r) &=&-{\frac { \left(n-2 \right)\left( { \alpha}^{2}+1 \right)
^{2}{b}^{-2\beta}}{\left( { \alpha}^{2}-1 \right)
\left(n+{\alpha}^{2}-2 \right)
}}{r}^{2\beta}-\frac{m}{r^{(n-1)(1-\beta )-1}}+\frac{2\Lambda
\left( {\alpha}^{2}+1 \right) ^{2}{b}^{2
\beta}}{(n-1)(\alpha^{2}-n )}r^{2(1-\beta)} \nonumber \\
&& +\frac{2(\alpha^2+1)^{2}b^{-2(n-2)\beta}q^{2}}{(n-1)(\alpha
^{2}+n-2)r^{2(n-2)(1-\beta )}} - \frac{(1+\alpha
^{2})^{2}b^{-2(2n-3)\beta }q^{4}}{\gamma (n-1)(\alpha
^{2}+3n-4)r^{2(2n-3)(1-\beta )}}
\end{eqnarray}
Note that in the limit $\gamma \rightarrow \infty$ and $\alpha
=\beta= 0$, it has the form of Reissner-Nordstrom (A)dS black hole
\cite{emparan}. The last term in the right hand side of the above
expression is the leading Born-Infeld correction to the RN(A)dS
black hole with dilaton field in the large $\gamma$ limit
\cite{CHM}.
\subsection*{Properties of the solutions}

\begin{figure}[tbp]
\epsfxsize=7cm \centerline{\epsffile{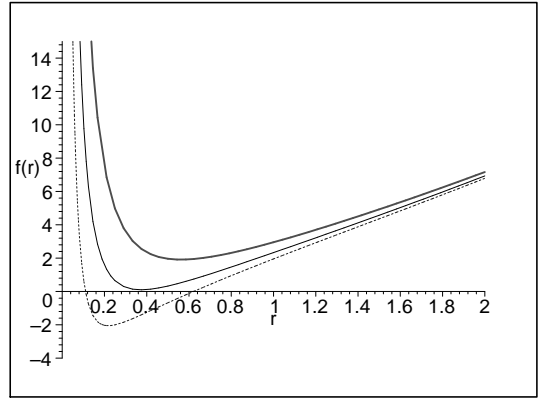}} \caption{The
function $f(r)$ versus $r$ for $\alpha=0.5$, $q=1$ and  $n=4$.
$m=1$ (bold), $m=1.6$ (continuous) and $m=2$ (dashed).}
\label{figureA}
\end{figure}

\begin{figure}[tbp]
\epsfxsize=7cm \centerline{\epsffile{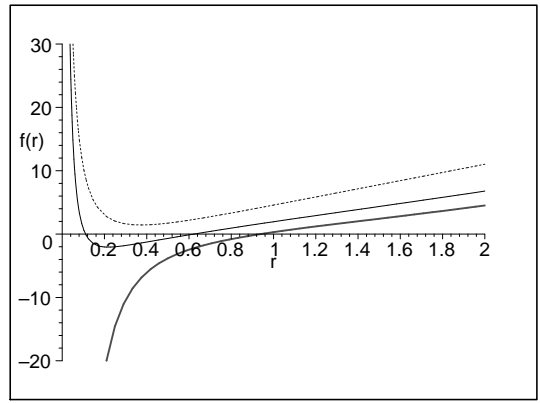}} \caption{The
function $f(r)$ versus $r$ for $q=1$, $m=2$ and $n=4$.
$\protect\alpha=0$ (bold), $\protect\alpha=0.5$ (continuous) and
$\protect\alpha=0.7$ (dashed).} \label{figureB}
\end{figure}
\begin{figure}[tbp]
\epsfxsize=7cm \centerline{\epsffile{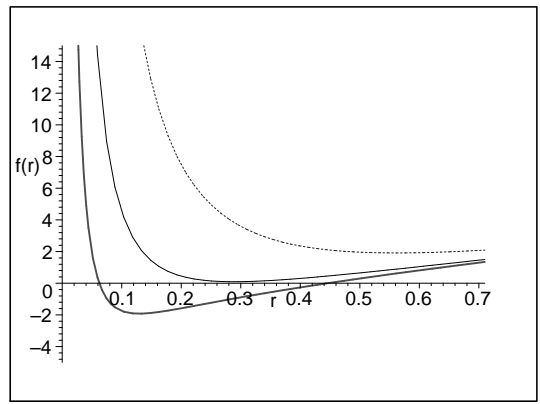}} \caption{The
function $f(r)$ versus $r$ for $m=1$ and $\alpha=0.5$ and $n=4$.
$q=0.55$ (bold), $q=0.66$ (continuous) and $q=1$ (dashed).}
\label{figureC}
\end{figure}
\begin{figure}[tbp]
\epsfxsize=7cm \centerline{\epsffile{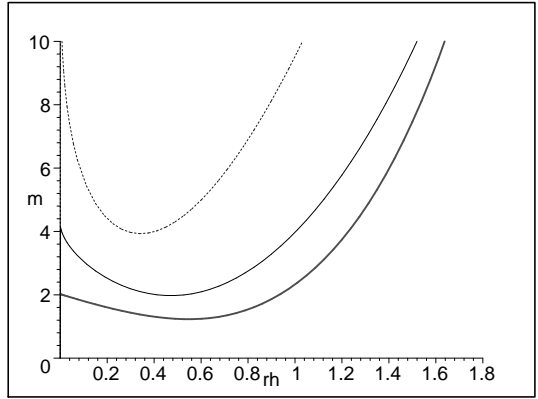}} \caption{The
function $m(r_{h})$ versus $r_{h}$ for $\gamma=2$ and $n=4$.
$\alpha=0$ (bold), $\alpha=0.5$ (continuous) and $\alpha=0.8$
(dashed).} \label{figureD}
\end{figure}
\begin{figure}[tbp]
\epsfxsize=7cm \centerline{\epsffile{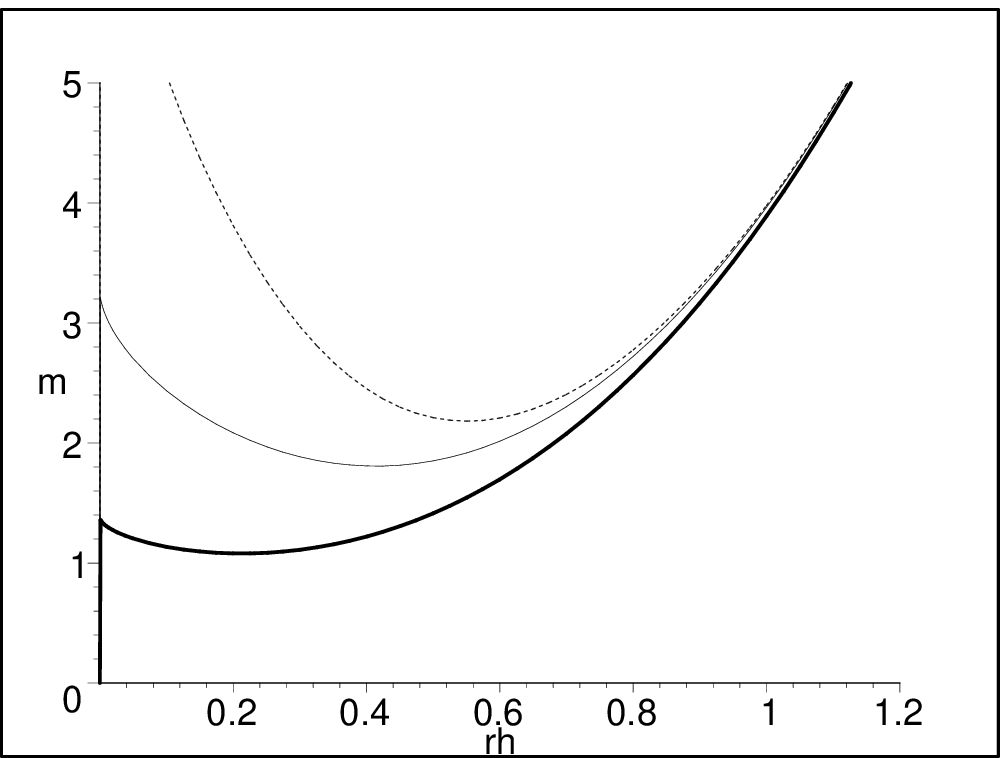}} \caption{The
function $m(r_{h})$ versus $r_{h}$ for  $\alpha=0.5$ and $n=4$.
$\gamma=0.1$ (bold), $\gamma=1$ (continuous) and $\gamma=10$
(dashed).} \label{figureE}
\end{figure}

Since the solution (\ref{f1}) can be obtained from (\ref{f2}) by
setting $\Lambda=0$, in the remains part of paper we focus on
(\ref{f2}). In order to study the general structure of these
solutions, we first look for the curvature singularities in the
presence of dilaton gravity. It is easy to show that in the
presence of dilaton field, the Kretschmann scalar $R_{\mu \nu
\lambda \kappa }R^{\mu \nu \lambda \kappa }$ diverges at $r=0$, it
is finite for $r\neq 0$ and goes to zero as $r\rightarrow \infty
$. Thus, there is an essential singularity located at $r=0$. The
spacetime is neither asymptotically flat nor (A)dS. It is notable
to mention that these solutions do not exist for the string case
where $\alpha=1$. As one can see from Eq. (\ref{f1}) and
(\ref{f2}), the solutions are ill-defined for $\alpha =\sqrt{n}$.
The cases with $\alpha>\sqrt{n}$ and $\alpha <\sqrt{n}$ should be
considered separately. In the first case where $\alpha >\sqrt{n}$,
 the spacetime has a cosmological horizon for positive
values of the mass parameter, despite the
sign of the cosmological constant $\Lambda $. In the second case where $%
\alpha <\sqrt{n}$,  there exist a cosmological horizon for
$\Lambda >0$, while there is no cosmological horizons if $\Lambda
<0$ . Indeed, in the latter case ($\alpha <\sqrt{n}$ and $\Lambda
<0$) the spacetimes associated with these solutions exhibit a
variety of possible casual structures
depending on the values of the metric parameters $\alpha $, $\gamma$, $m$, $q$, and $%
\Lambda $. One can obtain the casual structure by finding the
roots of $ f(r)=0$. Unfortunately, because of the nature of the
(\ref{f2}), it is not possible to find analytically the location
of the horizons. To achieve understanding of the nature of the
horizons, we first plot in figures
(\ref{figureA})-(\ref{figureC}), the function $f(r)$ for some
different values of mass parameter $m$ and dilaton coupling
$\alpha$. For simplicity we have kept fixed the other parameters
$l=1$, $b=1$, $\gamma=1$. We find from these figures that our
solutions can represent black hole, with inner and outer event
horizons, an extreme black hole or a naked singularity provided
the parameters of the solutions are chosen suitably. Then, we plot
in figures (\ref{figureD}) and (\ref{figureE}), the mass parameter
$m$ as a function of the horizon radius $r_{h}$ for some different
values of Born-Infeld parameter $\gamma$ and dilaton coupling
$\alpha$. Again we fixed $l=1$, $b=1$ and $q=1$, for simplicity.
It is easy to show that the mass parameter of the black hole can
be expressed in terms of the horizon radius as

\begin{eqnarray}\label{mass}
m(r_{h})&=&-{\frac { \left(n-2 \right)\left( { \alpha}^{2}+1
\right) ^{2}{b}^{-2\beta}{r_{h}}^{n-2+\beta(3-n)}}{\left( {
\alpha}^{2}-1 \right) \left(n+{\alpha}^{2}-2 \right)
}}+\frac{2\Lambda \left( {\alpha}^{2}+1 \right) ^{2}{b}^{2
\beta}r_{h}^{n(1-\beta)-\beta}}{(n-1)(\alpha^{2}-n
)}-\frac{2\gamma (\alpha ^{2}+1)^{2}b^{2\beta
}r_{h}^{n(1-\beta)-\beta}}{(n-1)(\alpha
^{2}-n)}\nonumber\\
&&\times
\left(1-\text{}_{2}F_{1}\left( \left[ -\frac{1}{2},\frac{\alpha ^{2}-n%
}{2n-2}\right] ,\left[ \frac{\alpha ^{2}+n-2}{2n-2}\right] ,\frac{-2q^{2}}{%
\gamma b^{2\beta (n-1)}r_{h}^{2(n-1)(1-\beta )}}\right) \right).
\end{eqnarray}
Figures (\ref{figureD}) and (\ref{figureE}), show that for a given
value of $\alpha$ or $\gamma$, the number of horizons depend on
the choice of the value of the mass parameter $m$. We see from
these figures that up to certain value of the mass parameter $m$,
there are two horizons, and as we decrease the $m$ further, the
two horizons meet. In this case we get extremal black hole.
Numerical calculations show that when this condition is satisfied,
the temperature of the black hole vanishes. To see this better we
obtain the temperature of the black hole on the horizon. The
Hawking temperature of the black hole on the outer horizon $r_{+}$
can be calculated using the relation
\begin{equation}
T_{+}=\frac{\kappa}{2\pi}= \frac{f^{\text{ }^{\prime
}}(r_{+})}{4\pi},
\end{equation}
where $\kappa$ is the surface gravity. Then, one can easily show
that
\begin{eqnarray}\label{Tem}
T_{+}&=&-\frac{(\alpha ^2+1)b^{2\beta}r_{+}^{1-2\beta}}{2\pi
(n-1)}\left(
\frac{(n-2)(n-1)b^{-4\beta}}{2(\alpha^2-1)}r_{+}^{4\beta-2}
+\left( \Lambda -\gamma\right)+\Gamma b^{(1-n)\beta}
r_{+}^{(1-n)(1-\beta)}\right)\nonumber\\
&=&\frac{(n-\alpha ^{2})m}{4\pi(\alpha
^{2}+1)}{r_{+}}^{(n-1)(\beta -1)}-\frac{(n-2)(\alpha
^2+1)b^{-2\beta}}{2\pi(\alpha ^2+n-2)}r_{+}^{2\beta-1}
-\frac{q^2(\alpha ^{2}+1)b^{2(2-n)\beta}}{\pi(\alpha
^2+n-2)}r_{+}^{2(2-n)(1-\beta)-1}\nonumber\\
&&\times \text{ }_{2}F_{1}\left( %
\left[ {\frac{1}{2},\frac{{n+\alpha }^{2}{-2}}{{2n-2}}}\right] ,\left[ {%
\frac{{3n+\alpha }^{2}{-4}}{{2n-2}}}\right]
,-\frac{2q^{2}b^{2\beta (1-n)}r_{+}^{2(n-1)(\beta-1 )}}{\gamma
}\right).
\end{eqnarray}
There is also an extreme value for the mass parameter in which the
temperature of the black hole is zero. It is a matter of
calculation to show that
\begin{eqnarray}
m_{\mathrm{ext}}&=&\frac{2(n-2)(\alpha
^2+1)^2b^{-2\beta}}{(n-\alpha ^{2})(\alpha
^2+n-2)}r_{\mathrm{ext}}^{(2-n)(\beta-1)+\beta} +\frac{4q^2(\alpha
^{2}+1)^2b^{2(2-n)\beta}}{(n-\alpha ^{2})(\alpha
^2+n-2)}r_{\mathrm{ext}}^{(3-n)(1-\beta)-1}\nonumber\\
&&\times \text{ }_{2}F_{1}\left( %
\left[ {\frac{1}{2},\frac{{n+\alpha }^{2}{-2}}{{2n-2}}}\right] ,\left[ {%
\frac{{3n+\alpha }^{2}{-4}}{{2n-2}}}\right]
,-\frac{2q^{2}b^{2\beta (1-n)}r_{\mathrm{ext}}^{2(n-1)(\beta-1
)}}{\gamma }\right).
\end{eqnarray}
Indeed the metric of Eqs. (\ref{metric}) and (\ref{f2}) has two
inner and outer horizons located at $r_{-}$ and $r_{+}$, provided
the mass parameter $ m $ is greater than $m_{\mathrm{ext}}$, an
extreme black hole in the case of $m=m_{\mathrm{ext}}$, and a
naked singularity if $m<m_{\mathrm{ext}}$. Notice that as we
increase $\alpha$ or $\gamma$, the extremal value of the mass,
$m_{\mathrm{ext}}$, increases too(see figs.
\ref{figureD}-\ref{figureE}).

\section{Thermodynamics of black hole} \label{Therm}
In this section we want to compute the conserved and
thermodynamics quantities of the EBId black hole. The metric
presented in this paper is not be asymptotically flat, therefore
we must use the quasilocal formalism to define the mass of the
solutions (see \cite{BY,Mann} for details). If we write the metric
of spherically symmetric spacetime in the form
\begin{equation}\label{metric2}
ds^2=-V(r)dt^2 + {dr^2\over V(r)}+ r^2d\Omega^2_{n-1},
\end{equation}
and the matter action contains no derivatives of the metric, then
the quasilocal mass is given by
\begin{equation}\label{QLM}
{\cal M} = \frac{n-1}{2}r^{n-2}{V(r)}^{1/2}\left( {V_{0}(r)}^{1/2}
- {V(r)}^{1/2}\right).
\end{equation}
Here $V_{0}(r)$ is an arbitrary function which determines the zero
of the energy for a background spacetime and $r$ is the radius of
the spacelike hypersurface boundary. When the spacetime is
asymptotically flat, the $ADM$ mass $M$ is the ${\cal M}$
determined in (\ref{QLM}) in the limit $r\rightarrow\infty$. If no
cosmological horizon is present, the large $r$ limit of
(\ref{QLM}), is used to determine the mass.  If a cosmological
horizon is present one can not take the large $r$ limit to
identify the quasilocal mass. However, one can still identify the
small mass parameter in the solution \cite{BY,Mann}. For the
solution under consideration, there is no cosmological horizon and
if we transform the metric in the form (\ref{metric2}), then we
obtain the mass of the black hole
\begin{equation}
{M}=\frac{b^{(n-1)\beta}(n-1) \omega _{n-1}m}{16\pi(\alpha^2+1)},\label{Mass}
\end{equation}
where $\omega _{n-1}$ is the volume of the unit $(n-1)$ sphere.
Black hole entropy typically satisfies the so called area law of
the entropy \cite{Beck}. This near universal law applies to almost
all kinds of black holes and black holes in Einstein gravity
\cite{hunt}. It is a matter of calculation to show that the
entropy of the black hole is
\begin{equation}
{S}=\frac{b^{(n-1)\beta}\omega _{n-1}r_{+}^{(n-1)(1-\beta
)}}{4},\label{Entropy}
\end{equation}
which shows that the area law holds for the black hole solutions
in dilaton gravity. Next, we calculate the electric charge of the
solutions. To determine the electric field we should consider the
projections of the electromagnetic field tensor on spacial
hypersurfaces. The normal to such hypersurfaces is
\begin{equation}
u^{0}=\frac{1}{N},\text{ \ }u^{r}=0,\text{ \
}u^{i}=-\frac{V^{i}}{N},
\end{equation}
where $N$ and $V^{i}$ are the lapse function and shift vector.
Then the electric field is $E^{\mu }=g^{\mu \rho
}e^{\frac{-4\alpha \phi }{n-1}}F_{\rho \nu }u^{\nu }$, and the
electric charge can be found by calculating the flux of the
electric field at infinity, yielding
\begin{equation}
{Q}=\frac{q\omega _{n-1}}{4\pi}.  \label{Charge}
\end{equation}
The electric potential $U$, measured at infinity with respect to
the horizon, is defined by
\begin{equation}
U=A_{\mu }\chi ^{\mu }\left| _{r\rightarrow \infty }-A_{\mu }\chi
^{\mu }\right| _{r=r_{+}},  \label{Pot}
\end{equation}
where $\chi=\partial_{t}$ is the null generator of the horizon.
One can easily show that the gauge potential $A_{t }$
corresponding to the electromagnetic field (\ref{Ftr}) can be
written as
\begin{eqnarray}\label{vectorpot}
A_{t}&=&\frac{qb^{(3-n)\beta }}{\Upsilon r^{\Upsilon }}\text{ }_{2}F_{1}\left( %
\left[ {\frac{1}{2},\frac{{n+\alpha }^{2}{-2}}{{2n-2}}}\right] ,\left[ {%
\frac{{3n+\alpha }^{2}{-4}}{{2n-2}}}\right] ,-\frac{2q^{2}}{\gamma
b^{2\beta (n-1)}r^{2(n-1)(1-\beta )}}\right),
\end{eqnarray}
where $\Upsilon =(n-3)(1-\beta )+1$. Therefore, the electric
potential may be obtained as
\begin{equation}
U=\frac{qb^{(3-n)\beta }}{ \Upsilon{r_{+}}^{\Upsilon }}\text{ }%
_{2}F_{1}\left( \left[ {\frac{1}{2},\frac{{n+\alpha }^{2}{-2}}{{2(n-1)}}}%
\right] ,\left[ {\frac{{3n+\alpha }^{2}{-4}}{{2(n-1)}}}\right] ,-\frac{2q^{2}%
}{\gamma b^{2\beta (n-1)}r_{+}^{2(n-1)(1-\beta )}}\right).
\label{Pot}
\end{equation}

Then, we consider the first law of thermodynamics for the black
hole. In order to do this, we obtain the mass $M$ as a function of
extensive quantities $S$, and $Q$. Using the expression for the
mass, the entropy and the charge given in Eqs. (\ref{Mass}),
(\ref{Entropy}) and (\ref{Charge}) and the fact that $f(r_{+})=0$,
one can obtain a Smarr-type formula as
\begin{eqnarray}
M(S,Q)&=&\frac{-(n-1)(n-2)(\alpha^2+1)b^{-\alpha^2}
{\left(4S\right)}^{\frac{\alpha^2+n-2}{n-1}}}{16\pi(\alpha^2-1)(\alpha^2+n-2)}
+\frac{(\alpha^2+1)(\Lambda-\gamma)b^{\alpha^2}{\left(4S\right)}^{\frac{n-\alpha^2}{n-1}}}{8\pi(\alpha^2-n)}
+\nonumber\\
&&\frac{(n-1)(\alpha^2+1)b^{\alpha^2}{\left(4S\right)}^{\frac{n-\alpha^2}{n-1}}}
{8\pi(\alpha^2-n)}\text{ }_{2}F_{1}\left( \left[ -\frac{1}{2},\frac{\alpha ^{2}-n%
}{2n-2}\right] ,\left[ \frac{\alpha ^{2}+n-2}{2n-2}\right] ,\frac{-2\pi^2Q^{2}}{%
\gamma S^2}\right) \label{Msmar}
\end{eqnarray}
One may then regard the parameters $S$, and $Q$ as a complete set
of extensive parameters for the mass $M(S,Q)$ and define the
intensive parameters conjugate to $S$ and $Q$. These quantities
are the temperature and the electric potential
\begin{equation}
T=\left( \frac{\partial M}{\partial S}\right) _{Q},\ \ U=\left( \frac{\partial M%
}{\partial Q}\right) _{S}.  \label{Dsmar}
\end{equation}
Numerical calculations show that the intensive quantities
calculated by Eq. (\ref{Dsmar}) coincide with Eqs. (\ref{Tem}) and
(\ref{Pot}). Thus, these thermodynamics quantities satisfy the
first law of thermodynamics
\begin{equation}
dM = TdS+Ud{Q}.
\end{equation}

\section{Stability in the canonical ensemble}\label{stab}
\begin{figure}[tbp]
\epsfxsize=7cm \centerline{\epsffile{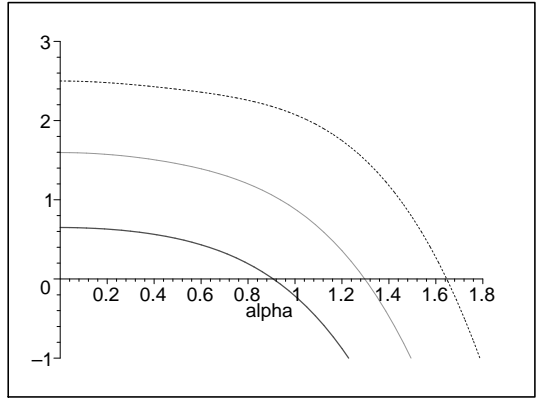}}
\caption{$(\partial ^{2}M/\partial S^{2})_{Q}$ versus
$\protect\alpha $ for  $q=1$, $\gamma=1$ and $n=5$. $q=0.5$
(bold), $q=1$ (continuous) and $q=1.5$ (dashed).} \label{Figure1}
\end{figure}

\begin{figure}[tbp]
\epsfxsize=7cm \centerline{\epsffile{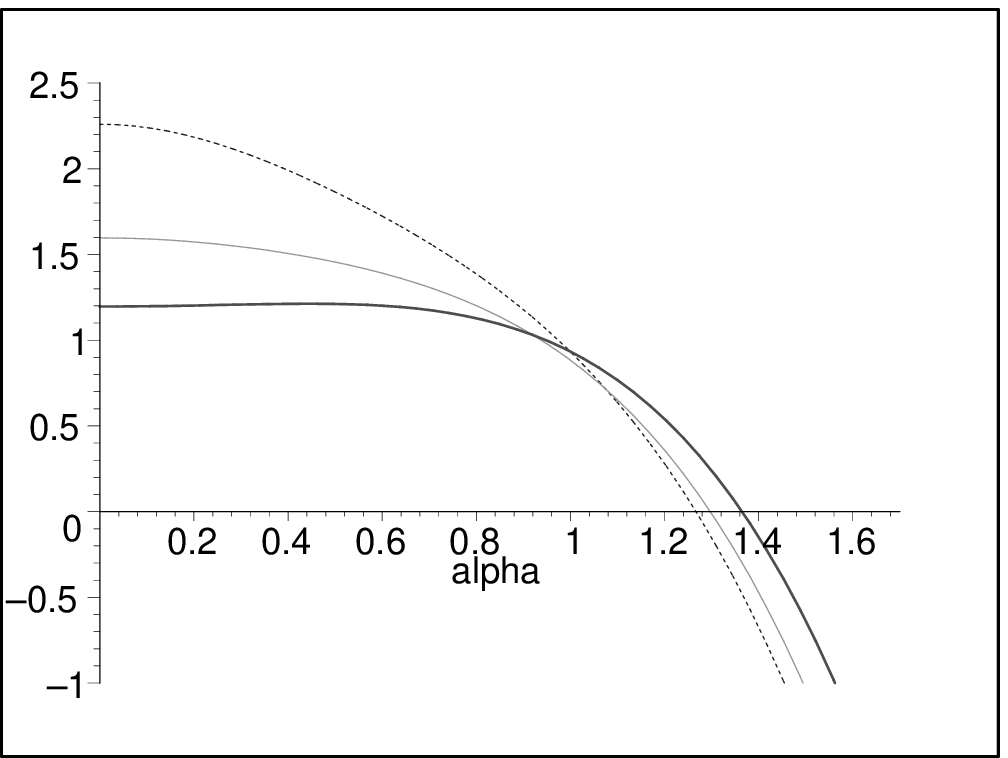}}
\caption{$(\partial ^{2}M/\partial S^{2})_{Q}$ versus
$\protect\alpha $ for $q =1$ and $\gamma=1$. $n=4$ (bold), $n=5$
(continuous) and $n=6$ (dashed).} \label{Figure2}
\end{figure}

\begin{figure}[tbp]
\epsfxsize=7cm \centerline{\epsffile{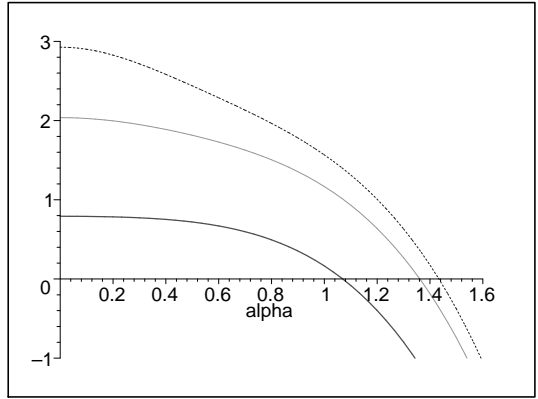}}
\caption{$(\partial ^{2}M/\partial S^{2})_{Q}$ versus
$\protect\alpha $ for  $n=5$ and $q=1$. $\gamma=0.2$ (bold),
$\gamma=2$ (continuous) and $\gamma=12$ (dashed).} \label{Figure3}
\end{figure}

\begin{figure}[tbp]
\epsfxsize=7cm \centerline{\epsffile{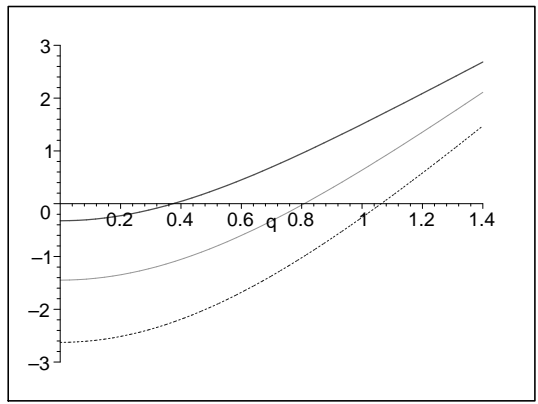}}
\caption{$(\partial ^{2}M/\partial S^{2})_{Q}$ versus $q$ for ,
$n=5$ and $\gamma=2$. $\protect\alpha=0.8$ (bold),
$\protect\alpha=1.2$ (continuous) and
$\protect\alpha=\protect\sqrt{2}$ (dashed).} \label{Figure4}
\end{figure}

\begin{figure}[tbp]
\epsfxsize=7cm \centerline{\epsffile{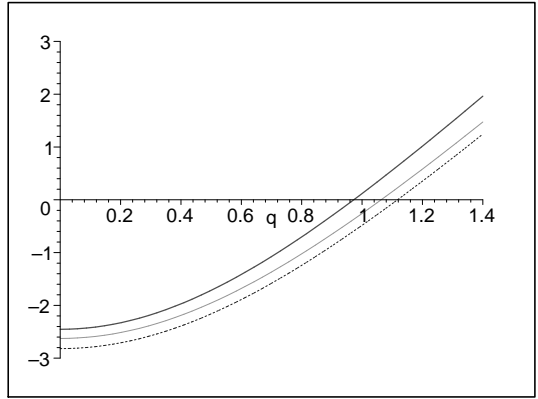}}
\caption{$(\partial ^{2}M/\partial S^{2})_{Q}$ versus $q $ for
$\gamma=2$ and $\protect\alpha=\protect\sqrt{2}$. $n=4$ (bold),
$n=5$ (continuous) and $n=6$ (dashed).} \label{Figure5}
\end{figure}`

\begin{figure}[tbp]
\epsfxsize=7cm \centerline{\epsffile{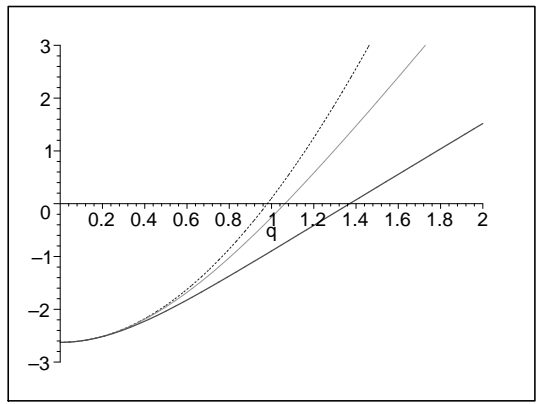}}
\caption{$(\partial ^{2}M/\partial S^{2})_{Q}$ versus $q $ for
$n=5$ and $\protect\alpha=\protect\sqrt{2}$. $\gamma=0.5$ (bold),
$\gamma=2$ (continuous) and $\gamma=12$ (dashed).} \label{Figure6}
\end{figure}

\begin{figure}[tbp]
\epsfxsize=7cm \centerline{\epsffile{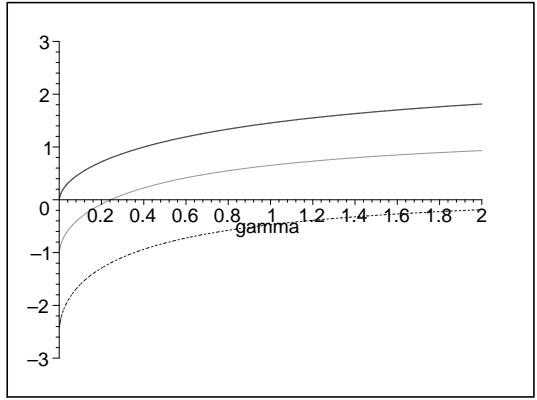}}
\caption{$(\partial ^{2}M/\partial S^{2})_{Q}$ versus $\gamma $
for $n=5$, $q=1$. $\protect\alpha=0.5$ (bold),
$\protect\alpha=1.1$ (continuous) and $\protect\alpha=1.4$
(dashed).} \label{Figure7}
\end{figure}

\begin{figure}[tbp]
\epsfxsize=7cm \centerline{\epsffile{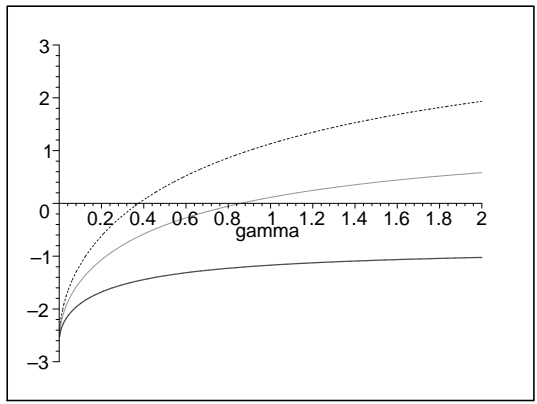}}
\caption{$(\partial ^{2}M/\partial S^{2})_{Q}$ versus
$\protect\gamma $ for $n=5$ and $\protect\alpha=\protect\sqrt{2}$.
$q=0.8$ (bold), $q=1.2$ (continuous) and $q=1.5$ (dashed).}
\label{Figure8}
\end{figure}

\begin{figure}[tbp]
\epsfxsize=7cm \centerline{\epsffile{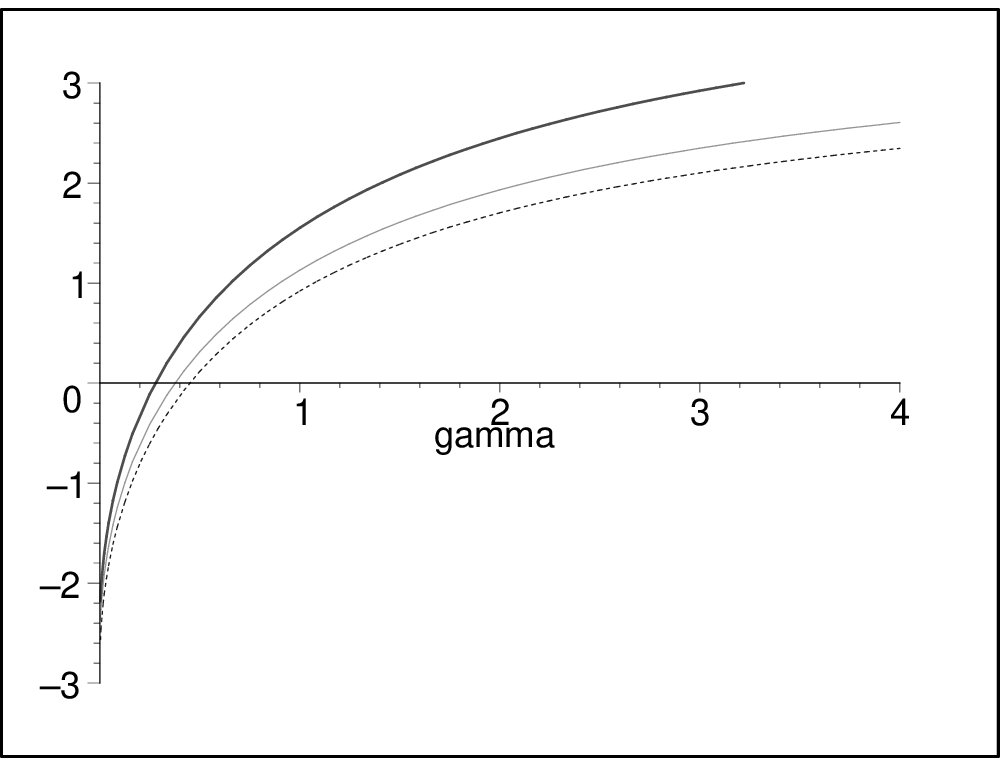}}
\caption{$(\partial ^{2}M/\partial S^{2})_{Q}$ versus $\gamma $
for  $q=1.5$ and $\protect\alpha=\protect\sqrt{2}$. $n=4$(bold),
$n=5$ (continues) and $n=6$ (dashed).} \label{Figure9}
\end{figure}
\begin{figure}[tbp]
\epsfxsize=7cm \centerline{\epsffile{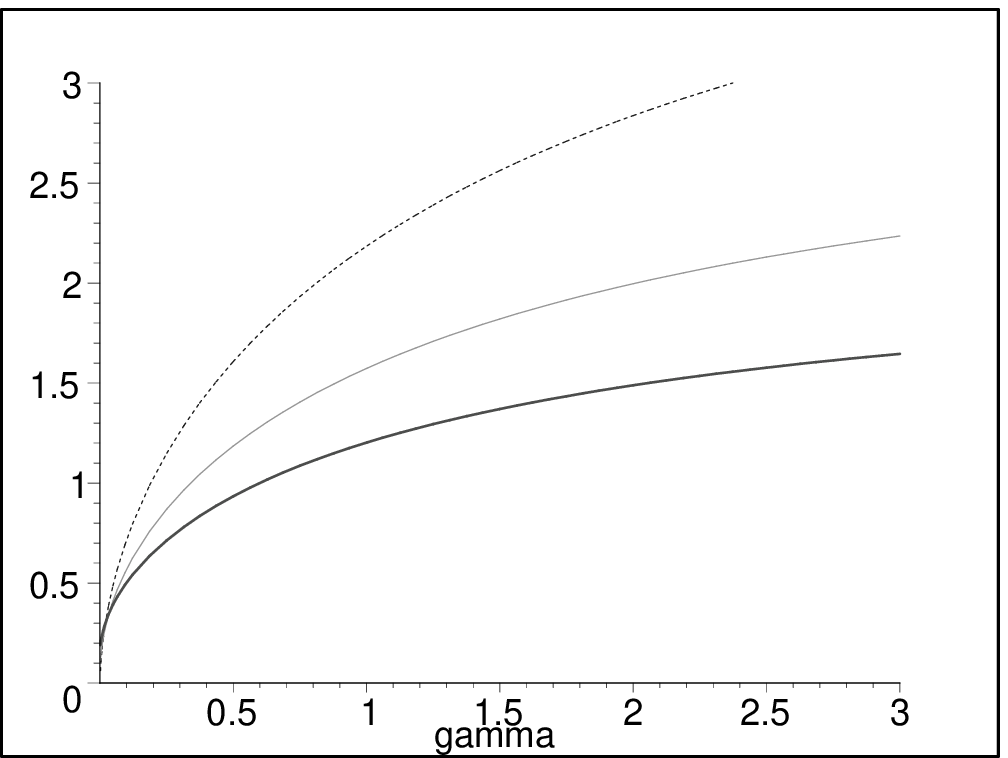}}
\caption{$(\partial ^{2}M/\partial S^{2})_{Q}$ versus $\gamma $
for $q=1$ and $\protect\alpha=0.2$. $n=4$(bold), $n=5$
(continuous) and $n=6$ (dashed).} \label{Figure10}
\end{figure}

Finally, we would like to study the stability of the EBId black
hole we have just found. Specially, we want to investigate the
effect of dilaton on the stability of the solutions. The stability
of a thermodynamic system with respect to small variations of the
thermodynamic coordinates is usually performed by analyzing the
behavior of the entropy $ S(M,Q)$ around the equilibrium. The
local stability in any ensemble requires that $S(M,Q)$ be a convex
function of the extensive variables or its Legendre transformation
must be a concave function of the intensive variables. The
stability can also be studied by the behavior of the energy
$M(S,Q)$ which should be a convex function of its extensive
variable. Thus, the local stability can in principle be carried
out by finding the determinant of the Hessian matrix of $M(S,Q)$
with respect to its extensive variables $X_{i}$, $
\mathbf{H}_{X_{i}X_{j}}^{M}=[\partial ^{2}M/\partial X_{i}\partial
X_{j}]$ \cite{Cal2,Gub}. In our case the mass $M$ is a function of
entropy and charge. The number of thermodynamic variables depends
on the ensemble that is used. In the canonical ensemble, the
charge is a fixed parameter and therefore the positivity of the
$(\partial ^{2}M/\partial S^{2})_{Q}$ is sufficient to ensure
local stability. Numerical calculations show that the black hole
solutions are stable independent of the value of the charge and
Born-Infeld parameters $q$ and $\gamma $, in any dimensions if
$\alpha < \alpha_{\max }$, while for $\alpha > \alpha_{\max }$ the
system has an unstable phase (see figs.
\ref{Figure1}-\ref{Figure3}). Notice that again we have kept fixed
the other parameters $l=1$, $b=1$, $r_{+}=.08$ in the figures
\ref{Figure1}-\ref{Figure10}. On the other hand, figures
\ref{Figure4}-\ref{Figure6} show that there is always a low limit
for electric charge $q_{\min}$\,  for which the system is
thermally stable provided $q > q_{\min }$. It is worth noting that
$\alpha _{\mathrm{\max }}$ and $q_{\min}$ depend on the
Born-Infeld parameter $\gamma $, and the dimensionality of space
time.  In figures \ref{Figure7}-\ref{Figure9} we plot $(\partial
^{2}M/\partial S^{2})_{Q}$ versus Born-Infeld parameter $\gamma$
for different values of the parameters $q$, $ \alpha$ and $n$.
These figures show that, there is always a low limit for
Born-Infeld parameter $\gamma _{\min }$, for which $(\partial
^{2}M/\partial S^{2})_{ Q}$ is positive provided $\gamma >\gamma
_{\min }$. It is notable to mention that $\gamma _{\min }$
decreases with increasing $q$ and increases with increasing
$\alpha$ and $n$. Finally, in figure \ref{Figure10} we show more
explicitly  the thermal stability of  the black hole solutions for
small value of dilaton coupling $\alpha$, in any dimensions for a
fixed value of the charge parameter $q$ and arbitrary Born-Infeld
parameter $\gamma$.

\section{Closing Remarks}

To sum up, we presented the $(n+1)$-dimensional
Einstein-Born-Infeld action coupled to a dilaton field and
obtained the equations of motion by varying this action with
respect to the gravitational field $g_{\mu \nu }$, the dilaton
field $\Phi $ and the gauge field $A_{\mu}$. Then, we constructed
a new class of charged, black hole solutions to
$(n+1)$-dimensional $(n\geq3)$ Einstein-Born-Infeld-dilaton theory
with one and two Liouville-type potentials and investigated their
properties. These solutions are neither asymptotically flat nor
(anti)-de Sitter. These solutions do not exist for the string case
where $\alpha=1$. In the presence of Liouville-type potential, we
obtained exact solutions provided $\alpha \neq \sqrt{n}$. In the
particular case $\gamma \longrightarrow \infty $, these solutions
reduce to the $(n+1)$-dimensional Einstein-Maxwell-dilaton black
hole solutions given in Ref. \cite{CHM}, while in the absence of a
non-trivial dilaton ($\alpha =\beta=0 $), the above solutions
reduce to the $(n+1)$-dimensional Born-Infeld black hole in the
presence of a cosmological constant presented in \cite{Dey,Cai2}.
We found that these solutions can represent black holes, with
inner and outer event horizons, an extreme black hole or a naked
singularity provided the parameters of the solutions are chosen
suitably. We also computed temperature, mass, entropy, charge and
electric potential of the black hole solutions and found that
these quantities satisfy the first law of thermodynamics. We
showed that these thermodynamic quantities are independent of the
Born-Infeld parameter $\gamma $. We found a Smarr-type formula and
performed a stability analysis in canonical ensemble by
considering $ (\partial ^{2}M/\partial S^{2})_{Q}$ for the charged
black hole solutions in $(n+1)$ dimensions and showed that there
is no Hawking-Page phase transition in spite of charge of the
black hole provided $\alpha \leq \alpha_{\max }$, independent of
the values of the charge and Born-Infeld parameter, $q$, $\gamma $
and the dimensionality of the space time, while the solutions have
an unstable phase for $\alpha \leq \alpha_{\max }$. We found that
there is always a low limit for electric charge $q_{\min}$\, for
which $(\partial ^{2}M/\partial S^{2})_{Q}$\ is positive, provided
$q>q_{\min }$. It is worth to note that $\alpha _{\mathrm{\max }}$
and $q_{\min}$ depend on the Born-Infeld parameter $\gamma $, and
the dimensionality of space time. On the other hand, we found a
lower limit for the Born-Infeld parameter $\gamma _{\min }$, for
which $(\partial ^{2}M/\partial S^{2})_{Q}$ is positive provided
$\gamma
>\gamma _{\min }$. It is notable to mention that $\gamma _{\min }$
decreases with increasing $q$ and increases with increasing
$\alpha$ and $n$. Note that the $(n+1)$-dimensional black hole
solutions obtained here are static. Thus, it would be interesting
if one could construct rotating black hole solutions in $(n+1)$
dimensions in the presence of dilaton and Born-Infeld fields.

\acknowledgments{This work has been supported in part by Shiraz
University.}

\end{document}